\documentclass[conference]{IEEEtran}
\IEEEoverridecommandlockouts
\usepackage[hidelinks]{hyperref}
\usepackage{cite}
\usepackage{amsmath,amssymb,amsfonts}
\usepackage{algorithmic}
\usepackage{graphicx}
\usepackage{textcomp}
\usepackage{xcolor}
\usepackage{multirow}
\usepackage[defaultlines=2,all]{nowidow}
\def\BibTeX{{\rm B\kern-.05em{\sc i\kern-.025em b}\kern-.08em
    T\kern-.1667em\lower.7ex\hbox{E}\kern-.125emX}}

\newcommand\copyrighttext{
  \begin{minipage}{\textwidth}\textcopyright 2022 IEEE. Personal use of this material is permitted. Permission from IEEE must be obtained for all other uses, in any current or future media, including reprinting/republishing this material for advertising or promotional purposes, creating new collective works, for resale or redistribution to servers or lists, or reuse of any copyrighted component of this work in other works.\end{minipage} }

\begin{document}
\copyrighttext

\title{HLS-based Optimization of Tau Triggering Algorithm for LHC: a case study\\
\thanks{This work was supported in part by the European Union through European Social Fund in the frames of the "Information and Communication Technologies (ICT) programme" ("ITA-IoIT" topic) and by the Estonian Research Council grant PUT PRG1467 "CRASHLESS".}
}

\author{\IEEEauthorblockN{Natalia Cherezova\IEEEauthorrefmark{1}, 
Dmitri Mihhailov\IEEEauthorrefmark{1},
Sergei Devadze\IEEEauthorrefmark{1},
Artur Jutman\IEEEauthorrefmark{2}}
\IEEEauthorblockA{\IEEEauthorrefmark{1}Department of Computer Systems, Tallinn University of Technology, Tallinn, Estonia}
\IEEEauthorblockA{\IEEEauthorrefmark{2}Testonica lab, Tallinn, Estonia}
\IEEEauthorblockA{\{natalia.cherezova, dmitri.mihhailov, sergei.devadze\}@taltech.ee, artur@testonica.com}
}

\IEEEoverridecommandlockouts
\IEEEpubid{\makebox[\columnwidth]{978-1-6654-6295-2/22/\$31.00~\copyright2022 IEEE \hfill} \hspace{\columnsep}\makebox[\columnwidth]{ }}

\maketitle

\IEEEpubidadjcol

\begin{abstract}
With the current increase in the data produced by the Large Hadron Collider (LHC) at CERN, it becomes important to process this data in a corresponding manner. To begin with, to efficiently select events that contain relevant information from a massive flow of data. This is the task of the tau lepton decay triggering algorithm. The implementation is based on the High-Level Synthesis (HLS) approach that allows generating a hardware description of the design from the algorithm written in a high-level programming language like C++. HLS tools are intended to decrease the time and complexity of hardware design development, however, their capabilities are limited. The development of an efficient application requires substantial knowledge of the hardware design and HLS specifics. This paper presents the optimizations introduced to the algorithm that improved latency and area and more importantly solved the problems with the routing, making it possible to implement the algorithm on the FPGA fabric.
\end{abstract}

\begin{IEEEkeywords}
HLS, algorithm optimization
\end{IEEEkeywords}

\section{Introduction}
With the current increase in the data produced by the Large Hadron Collider (LHC) at CERN, it becomes important to process this data in a corresponding manner. Firstly, to efficiently select events that contain relevant information from a massive flow of data. This is the main objective of the tau lepton decay triggering algorithm.

Tau lepton is an important particle for analysis of the physical processes happening in LHC, it "plays an important role in both precise measurement of Standard Model physics and search for physics beyond the Standard Model" \cite{Sakurai2014}. However, it has a short lifetime and short decay length and can be found and reconstructed only by its decay products. The tau triggering algorithm is designed to identify the events that have hadronically decaying tau leptons \cite{Sakurai2014}.

The algorithm consists of 3 major steps. In the first step, the event data is buffered, and the 16 objects (seeds) with the highest pT (transverse momentum) value are selected. Selected seeds and buffered data from the whole event are then passed to the select candidates step. The task of this step is to select up to 30 tau candidates from the neighborhood of each seed. Then, selected candidates are analyzed to find information that can be used to reconstruct tau objects.

The implementation of the algorithm is based on the High-Level Synthesis (HLS) approach that allows generating a hardware description from the algorithm written in a high-level programming language like C++. HLS tools are intended to decrease the time and complexity of hardware design development, which is especially useful in the case of compute- and data-intensive applications. However, the capabilities of HLS tools are limited. Proper optimization of the algorithm requires knowledge of the hardware design and HLS specifics.

The algorithm is developed using Vivado HLS targeting Xilinx Virtex UltraScale+ FPGA VCU118 Evaluation Kit.

The task of the case study was to optimize the critical parts of the algorithm in order to solve problems with timing, area, and routing. Considering the selected development approach, in order to optimize the algorithm two tasks should be solved: first, optimized design development and, second, its implementation in a correct way so that Vivado HLS can synthesize it accordingly.

The paper is organized in the following way. Section 2 presents the main concepts of high-level synthesis and its current state. Section 3 presents the sorting algorithm developed for the project. Section 4 provides an overview of the candidates selection task optimization. Conclusions are given in Section 5.

\section{High-level synthesis}
With the increased popularity of the FPGAs (Field-Programmable Gate Array) and heterogeneous systems combining processor units and programming logic on one platform, there arises the necessity to introduce new programming methods for the FPGAs that will make them more accessible to the broader public.

FPGAs have certain benefits over traditional computational platforms like CPU and GPU. Due to their natural parallel computing capabilities, they are faster than the CPU. While GPUs are capable of parallel computing as well, FPGAs are characterized by lower cost and much lower power consumption.

However, the complexity of programming that requires substantial knowledge of the underlying architecture and hardware specifics and increased time to develop the product makes FPGA platforms less accessible to the broader public. Using IP (Intellectual Property) cores, ready-to-use functional blocks, is one way to solve the problem; they allow building an application in a Lego-like fashion by combining needed blocks together. Nevertheless, not every functionality can be implemented using IPs. More complex algorithms require tailored solutions designed specifically for the application.

A better way to solve the problem is the High-Level Synthesis (HLS) tools that introduce a higher level of abstraction for the hardware programming, namely they allow developers to write algorithms in high-level languages like C and C++ that will be automatically translated by the tool into HDL specification considering characteristics of the selected board. This way, HLS tools allow non-hardware specialists to program FPGAs. Additionally, they increase the portability and maintainability of the application, code written in C/C++ can be easily re-synthesized for different target platforms, debugged, and changed in case there is a need. Since the input code for HLS describes the algorithm, what the program does, rather than how exactly it is implemented on the target platform.

However, HLS is not a panacea. It still requires domain-specific knowledge to write well-optimized code. HLS tools suggest different pragma directives for hardware-specific optimization of the algorithm. Nevertheless, those directives alone are not enough to synthesize optimal RTL implementation. They will not be able to improve the algorithm not suitable for the hardware implementation. In order to produce an efficient hardware description, the input code should be restructured accordingly. Licht et al. \cite{Licht2021} states that naive unoptimized HLS implementations show worse performance than naive software implementations. Matai et al. \cite{Matai2016} present a case study for insertion sort optimization, showing how properly written code can increase the performance of the algorithm, reducing the latency and resource usage. Restructured code shows the best result time- and area-wise compared to the software version of the algorithm optimized with different pragma directives.

Huang et al. \cite{Huang2020} compare the HLS-based development process to embedded systems development. Even though companies like STM and Arduino producing development boards provide great support for their products including frameworks for automated settings and board-specific libraries in order to make the programming of the devices easier, it still requires domain-specific knowledge to write efficient applications and solve arising problems.

The main difference is that high-level languages used in HLS for input code are not designed to describe hardware specifications, they are designed to describe software instructions executed sequentially, which introduces a challenge of describing hardware constructs with software languages.

Additionally, it requires the knowledge of how HLS tools themselves work, in what order they implement directives, how they represent the algorithm internally, and how they synthesize different constructs in order to restructure the design of the application in a way that will produce a desirable result.

Several works report that the quality of results (QoR) of the HLS generated hardware description is far behind that of manually written RTL design \cite{Huang2020, Licht2021, Liang2012, Rupnow2011, Lahti2019}. On the contrary, \cite{Cong2011a} shows an HLS-generated design that reduces resource usage by 11-13 \% compared to the design written by an RTL expert, proving that HLS can produce competitive results. The case study presented in \cite{ZhipengZhao2017} shows comparable results as well.

A serious limitation of HLS tools is that physical layout and routing estimation is difficult on the HLS level and therefore require again logic synthesis and implementation in order to identify whether synthesized design has routing problems and congestions. Reported congested areas should then be mapped to the original code, which is a problem on its own since the generated hardware description of the design differs drastically from the input code.

Nevertheless, the complex nature of the algorithm which is both compute-intensive and data-intensive calls for the HLS-based development process. And therefore, the aforementioned specifics of the HLS should be addressed and solved during the work on the project.

\subsection{Vivado HLS overview}

Vivado HLS \cite{VivadoHLS} is the commercial tool provided by Xilinx. It accepts code written in C, C++, and SystemC as input and synthesizes hardware description in VHDL, Verilog, and SystemC. The tool includes a variety of pragma directives for design optimization and libraries for hardware-specific features, such as arbitrary precision data types, Xilinx IP (Intellectual Property) functions for streams, shift registers, etc. Additionally, it provides automatic test-bench creation, C and RTL co-simulation, and support for floating-point and fixed-point arithmetics.

After synthesis, Vivado HLS creates a report describing the performance metrics of the generated design, including the maximum frequency of the design based on the longest combinational delay, latency of the design, initiation interval (number of clock cycles before the application can accept new input), number of utilized resources based on the number of resources available on the target platform, types of interfaces used for input and output signals. The same information is presented for every function and loop instantiated in the design.

In order to guarantee the synthesizability of the design, Vivado HLS does not accept some C/C++ language constructs, including recursion, dynamic memory allocation, function pointers, and operating system calls.

\section{Sorting}
The first step of the algorithm is dedicated to buffering the input data and selecting 16 seeds out of 144 candidates with the highest pT (transverse momentum) value. The candidates are sorted, and 16 elements from the top are saved into the Seeds array for further analysis.

Input data is a continuous stream of particle events. The event data includes data from 36 regions that come one region at a time. Every region contains 22 charged tracks, 13 photon tracks, and 10 tracks from neutral particles. First 4 tracks from every region form an array of seed candidates $(36\times4 = 144)$. Later an array of seed candidates is sorted to find the 16 best seeds based on their pT value since a high pT value can be an indication of decaying tau leptons.

\subsection{Original algorithm}

The original sorting algorithm was a hardware-optimized version of bubble sort that was taking too many resources causing problems with the routing of the whole algorithm. The task was to develop a new sorting algorithm that will significantly decrease the number of utilized resources and potentially decrease the latency of the first step since the total latency of the algorithm is very strict. The latency of the first step can be defined as $max(B, S)$, where $B$ is the latency of the buffering process and $S$ is the latency of the sorting. Buffering takes 56 clock cycles, the original sorting solution takes 57 clock cycles. Therefore, an optimized solution should take the same amount of clock cycles or less than the buffering.

In the beginning, it was decided to try some modifications of the original algorithm to explore possible optimizations without introducing a completely new sorting algorithm. Each seed candidate object is represented by a structure with 6 members. During the sorting, seed candidates are read and written numerous times. The idea was to use a smaller structure containing the pT value and index of the candidate for the sorting and then write the 16 best seeds using obtained indices. Another idea was instead of a smaller structure use a temporary array of $8+16$ bit integers that will hold both the index of the candidate (8 bits) and its pT value (16 bits). Both approaches only slightly reduced the resource usage of the function, did not reduce the latency and introduced the problem of using many multiplexers at the stage when the selected seeds should be written to the output array based on their indices in the seed candidates array. The results are presented in Table~\ref{tab:orig_sorting}.

\begin{table}[h!]
    \caption{Results of the original algorithm modifications}
    \begin{center}
    \begin{tabular}{|l|l|l|l|}
        \hline
        \textbf{Algorithm}  & \textbf{Latency, cycles}  & \textbf{FF}  & \textbf{LUT} \\
        \hline
        Original & 57 & 151,287 (6\%) & 262,690 (22\%) \\
        \hline
        Smaller struct & 58 & 132,448 (5\%) & 208,487 (17\%) \\
        \hline
        $8+16$ bit integer & 72 & 146,627 (6\%) & 238,428 (20\%) \\
        \hline
    \end{tabular}
    \label{tab:orig_sorting}
    \end{center}
\end{table}

Eventually, it was decided to try a different sorting algorithm. The algorithm should consider the streaming nature of incoming data, the fact that it comes in chunks of 4, and that only 16 elements out of 144 should be saved for further processing. The straightforward approach to use some hardware-optimized sorting architecture to sort all 144 elements would not be applicable, because even sorting algorithms with time complexity $O(n)$ would take much more clock cycles than desired.

\subsection{Streaming merge sort}

At first, it was decided to implement a merge sorter tree that at every level will leave only the 16 best elements, discarding the others.

On the first level, there are 16 arrays, 9 elements each $(16\times9 = 144)$. On the second level, they are merged into 8 arrays of size 16, and so on until there is only one array left with the best seeds. 16 arrays on the first level were sorted in an insertion sort manner: a new coming element would traverse through the array to find its place, then if needed elements greater than the new one will be shifted, and a new element inserted. To parallelize the process, each seed candidate from one region would be inserted into a different array. This way, a new element is inserted in the array on the first level every third cycle, giving the previous element 2 clock cycles to find its place, which was proved sufficient during the runs of the algorithm.

Three different variants of the merge sorter were tried: OUT arrays implemented as BRAMs, OUT arrays partitioned and implemented as separate registers, and finally OUT arrays implemented as FIFO streams. The results prove that streams are the best solution for the merge sorter as they decrease the latency of the merge stage. However, they have a certain drawback: all elements from the previous level that are not passed to the next level should anyway be read from the stream. The results are presented in Table~\ref{tab:merge_results}.

\begin{table}[h!]
    \caption{Results for merge sort implementations}
    \begin{center}
    \begin{tabular}{|l|l|l|l|l|}
        \hline
        \textbf{OUT arrays} & \textbf{Latency, c} & \textbf{BRAM} & \textbf{FF} & \textbf{LUT} \\
        \hline
        As BRAM & 104 & 48 & 35,333 (1\%) & 146,998 (12\%) \\
        \hline
        Partitioned & 84 & 0 & 63,552 (2\%) & 439,358 (37\%) \\
        \hline
        As streams & 72 & 0 & 41,117 (1\%) & 391,176 (33\%) \\
        \hline
    \end{tabular}
    \label{tab:merge_results}
    \end{center}
\end{table}

It can be seen that the implemented merge sort does not fit into defined latency and does not decrease the number of resources used and therefore does not satisfy the task description. The increased latency is attributed to the fact that it was required to divide the buffering process and the merge sorter between different functions. Buffering requires a pipeline directive in order to read new region data every clock cycle and start processing new event data every 36 cycles. Merge sorter though requires a dataflow directive to make all merge levels work in parallel, and Vivado~HLS cannot instantiate the dataflow region from pipelined function. Therefore, the merge sorter waits until the buffering stage is over and starts to work only after.

\subsection{Spatial insertion sort}

\begin{figure*}[h!]
    \centerline{\includegraphics[width=5.7in]{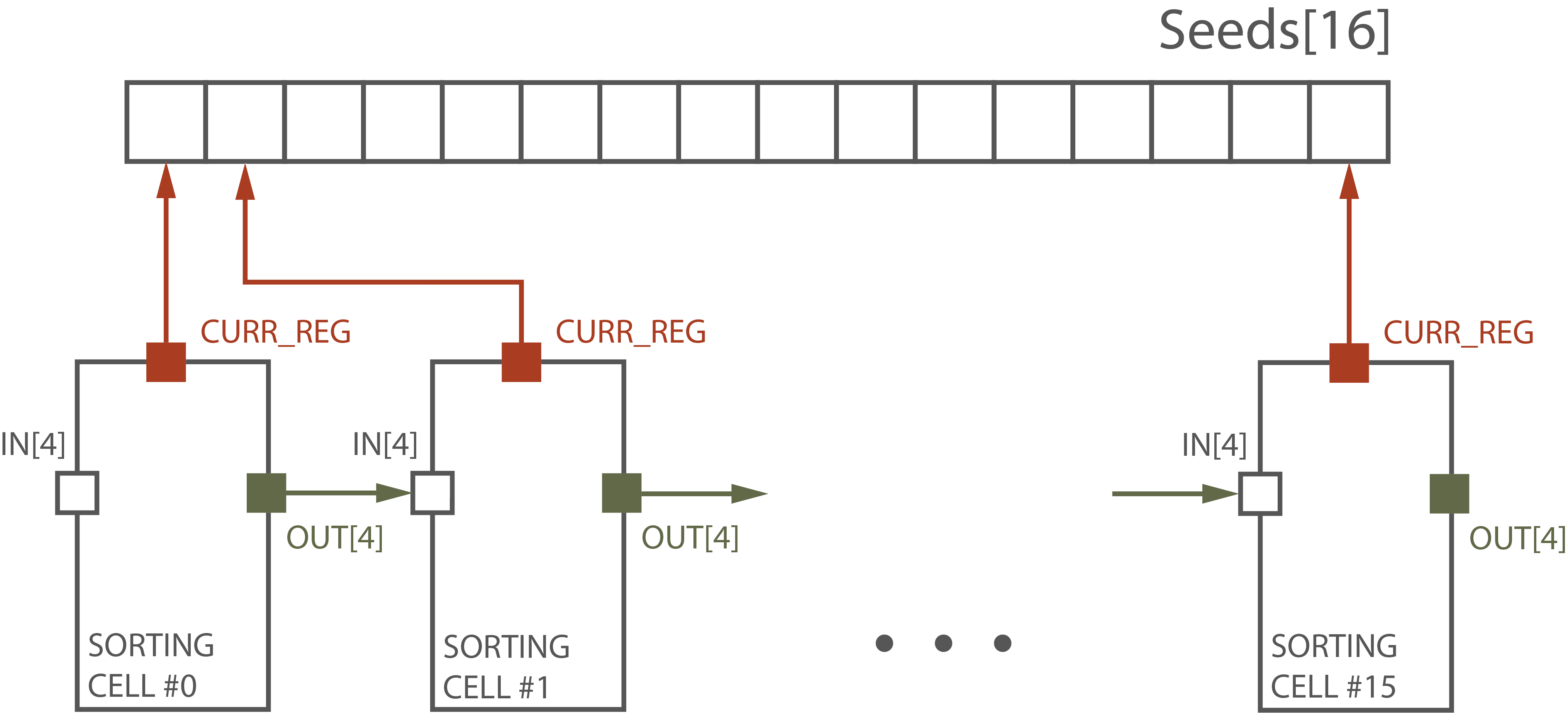}}
    \caption{Spatial sorter architecture}
    \label{fig:insertion}
\end{figure*}

The first algorithm developed that improved the timing and the resource usage was a modification of insertion sort. A spatial sorter was built that included 16 insertion cells. The sorting architecture is presented in Figure~\ref{fig:insertion}.

The sorting algorithm considers that 4 inputs from each region are already sorted and, therefore, 4 elements are used as an input for each insertion cell, unlike traditional implementation that has 1 new coming element as an input.

\begin{figure}[!h]
    \centerline{\includegraphics[width=3in]{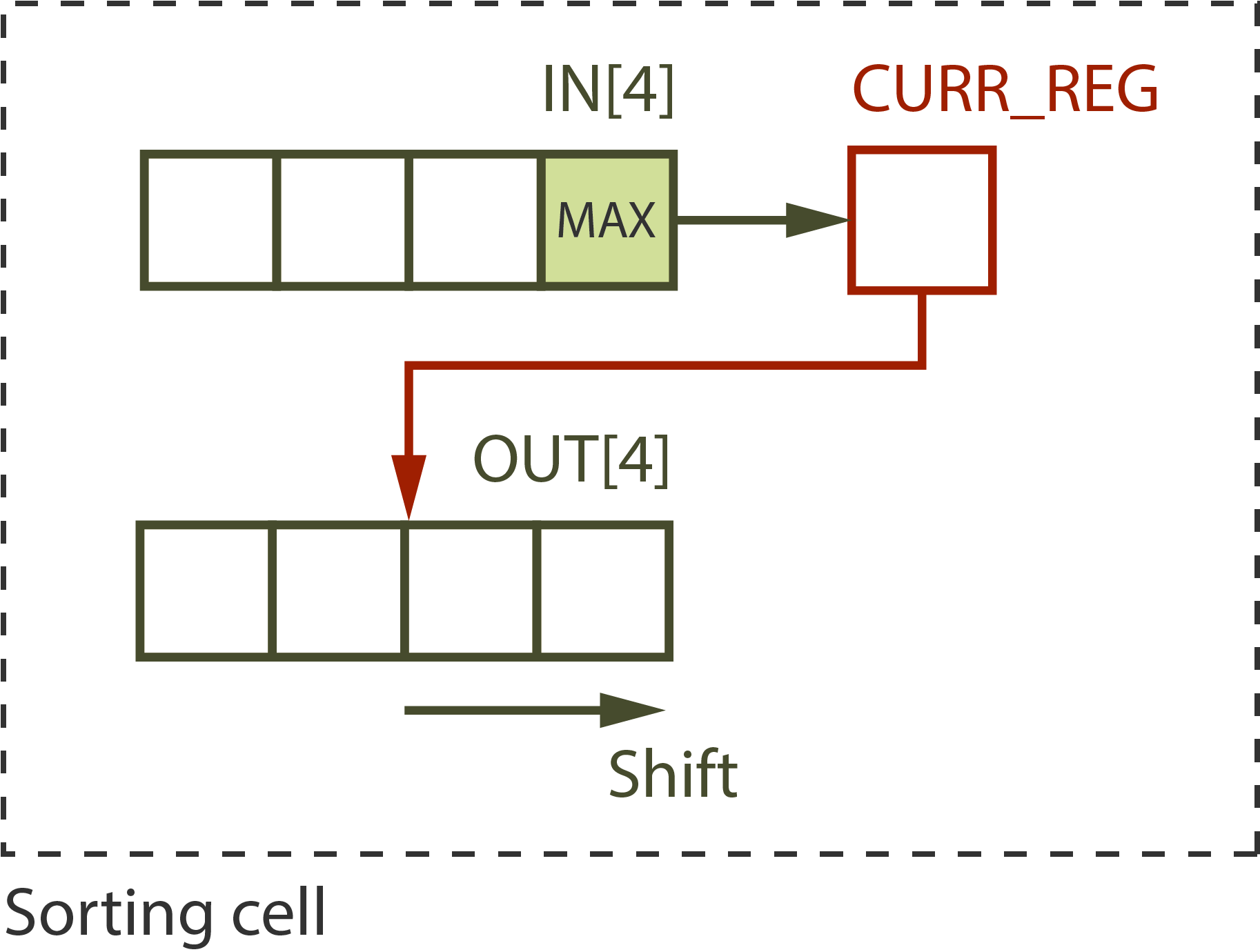}}
    \caption{Sorting cell architecture}
    \label{fig:insertion_cell}
\end{figure}

Each insertion cell works on one element of the Seeds array. First, the insertion cell gets 4 elements from the new region and compares it to the first element of the Seeds array. The biggest element is saved to the CURR\texttt{\char`_}REG variable. Since it is known that elements in the IN array are sorted, the biggest element is either CURR\texttt{\char`_}REG or IN[0]. If it is not CURR\texttt{\char`_}REG, IN[0] is saved into the CURR\texttt{\char`_}REG, the previous value of CURR\texttt{\char`_}REG is inserted in the appropriate place in the OUT array. The OUT array is then passed to the next sorting cell. The work of the insertion cell is presented in Figure~\ref{fig:insertion_cell}.

In order to make Vivado~HLS implement 16 different instantiations of the insertion cell function, so each of them will use its own static variable CURR\texttt{\char`_}REG, the function is defined as a template with an integer parameter that acts as an ID number. Otherwise, Vivado~HLS will generate the design, in which all insertion cells will share the same CURR\texttt{\char`_}REG variable.

Generally, the complexity of a spatial sorter is $O(2n)$, but in our case, the complexity is $O(n/4+m)$, where $n$ is the number of seed candidates and $m$ is the number of top seeds.

With this approach, sorting takes $36 + 1 + 16$ (53 cycles), because the first region elements are passed to the first sorting cell on the second cycle and the elements from the last region take 16 cycles to pass through all sorting cells. Since the latency of the first step function is determined by either buffering or sorting, depending on which operation takes more clock cycles, the total latency of the function with the described sorting algorithm is 56 clock cycles.

\begin{figure*}[htb]
    \centerline{\includegraphics[width=5.7in]{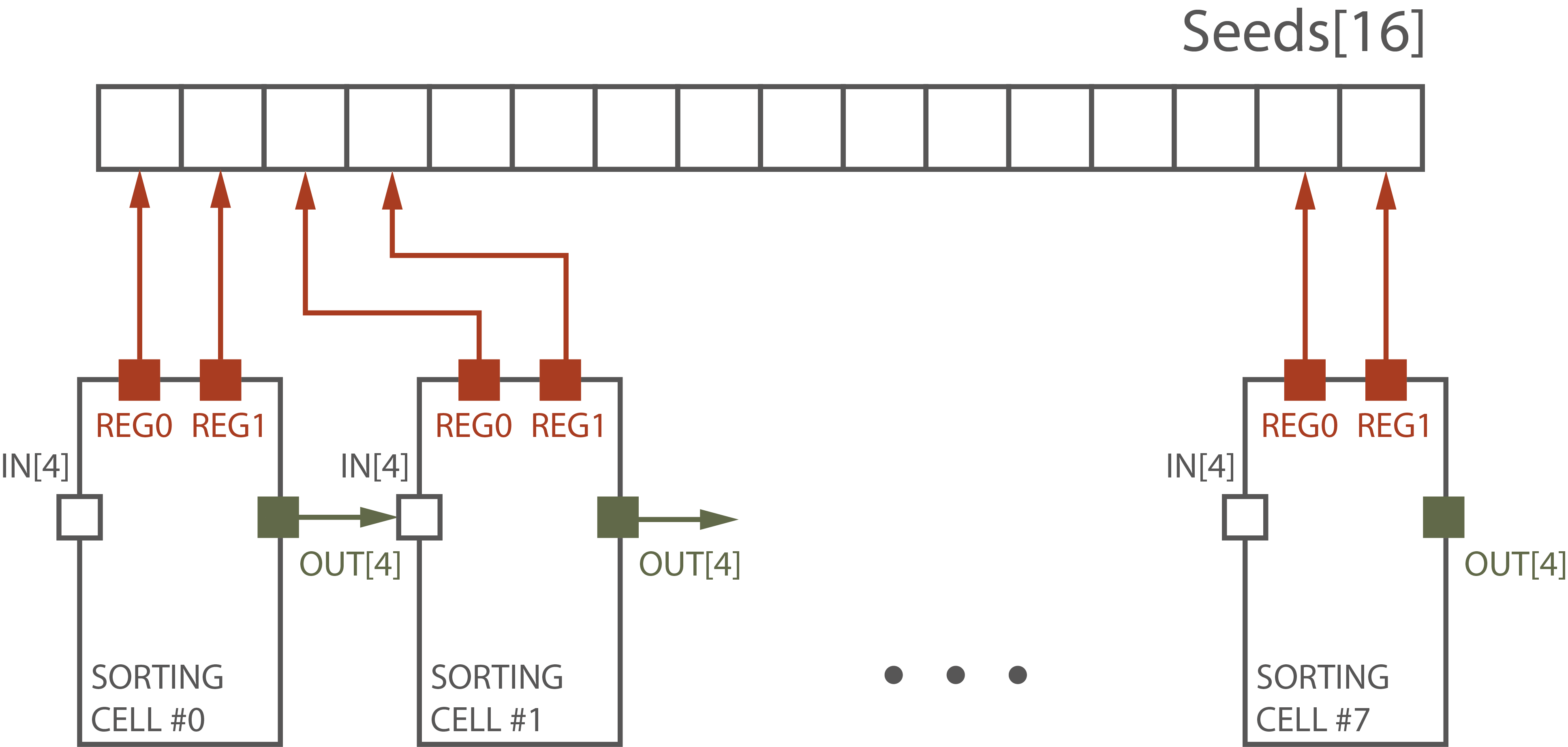}}
    \caption{Modified spatial sorter}
    \label{fig:sorter6}
\end{figure*}

It was possible to modify the created algorithm to decrease the timing of the sorting part even more. Instead of working on 1 element from the Seeds array, new sorting cells work on 2 elements from the Seeds array. Graphically, a modified sorter is presented in Figure~\ref{fig:sorter6}.

The architecture of the insertion cell has changed as well. The solution for the modified insertion cell was inspired by \cite{Kent2021} that presents a single-stage $N$ sorter based on a comparison counting matrix. The suggested $N$-sorter does not require calculating the rank of each element, it is built according to the equations derived from the comparison counting result.

Since it is known that the IN array is sorted and REG0 and REG1 are sorted, several simple rules were deducted to sort those elements. 6 elements should be sorted into 6 output positions: out1, out2, out3, out4, out5, and out6. out1 and out2 are saved into REG0 and REG1, and out3--out6 are saved into the OUT array.

The examples of the rules are presented below:
\begin{enumerate}
    \item Only two elements can go to the out1: REG0 or IN[0]. The biggest one goes to the out1.
    \item Four elements can go to the out2: REG0, REG1, IN[0], or IN[1]. If REG1 is bigger than IN[0], then it goes to the out2. If IN[0] is smaller than REG0 but greater than REG1, then it goes to the out2. If IN[1] is bigger than REG0, then it goes to the out2. Else, REG0 goes to the out2.
    \item The rules for the other output positions are depicted in the same way.
\end{enumerate}
The latency of the modified sorting algorithm is $36 + 1 + 8$ (45) cycles. The resource usage for the spatial sorter modifications is presented in Table~\ref{tab:insertion_results}.

\begin{table}[h!]
    \caption{Results for spatial insertion sorter versus original algorithm}
    \begin{center}
    \begin{tabular}{|l|l|l|l|}
        \hline
        \textbf{Algorithm} & \textbf{Latency, c} & \textbf{FF} & \textbf{LUT} \\
        \hline
        Original & 57 (57)* & 151,287 (6\%) & 262,690 (22\%) \\
        \hline
        Spatial sorter & 56 (53) & 104,749 (4\%) & 33,822 (2\%) \\
        \hline
        Modified spatial sorter & 56 (45) & 103,783 (4\%) & 22,987 (1\%) \\
        \hline
        \multicolumn{4}{l}{} \\
        \multicolumn{4}{l}{*~The latency of the sorting process is given in the brackets.}
    \end{tabular}
    \label{tab:insertion_results}
    \end{center}
\end{table}

It can be seen that a developed spatial sorter significantly decreases the usage of the resources and provide an opportunity to decrease the latency of the first step function if it will be possible to decrease the latency of the buffering process.

\section{Candidates selection}

During the second step of the algorithm, for each seed, the data from four neighboring regions (the region where the seed was found plus three adjacent ones based on the seed's location in the region) should be saved into candidate arrays. However, there was a problem with the way data from the selected regions was read from the input buffer and written to the candidate arrays. Regions were referred by their indices (from 0 to 35), but access through non-sequential indices created huge multiplexers and caused serious routing problems. A new representation of selected regions and a new way to write the data from the buffered arrays were required.

\begin{figure}[h!]
    \centerline{\includegraphics[width=1.5in]{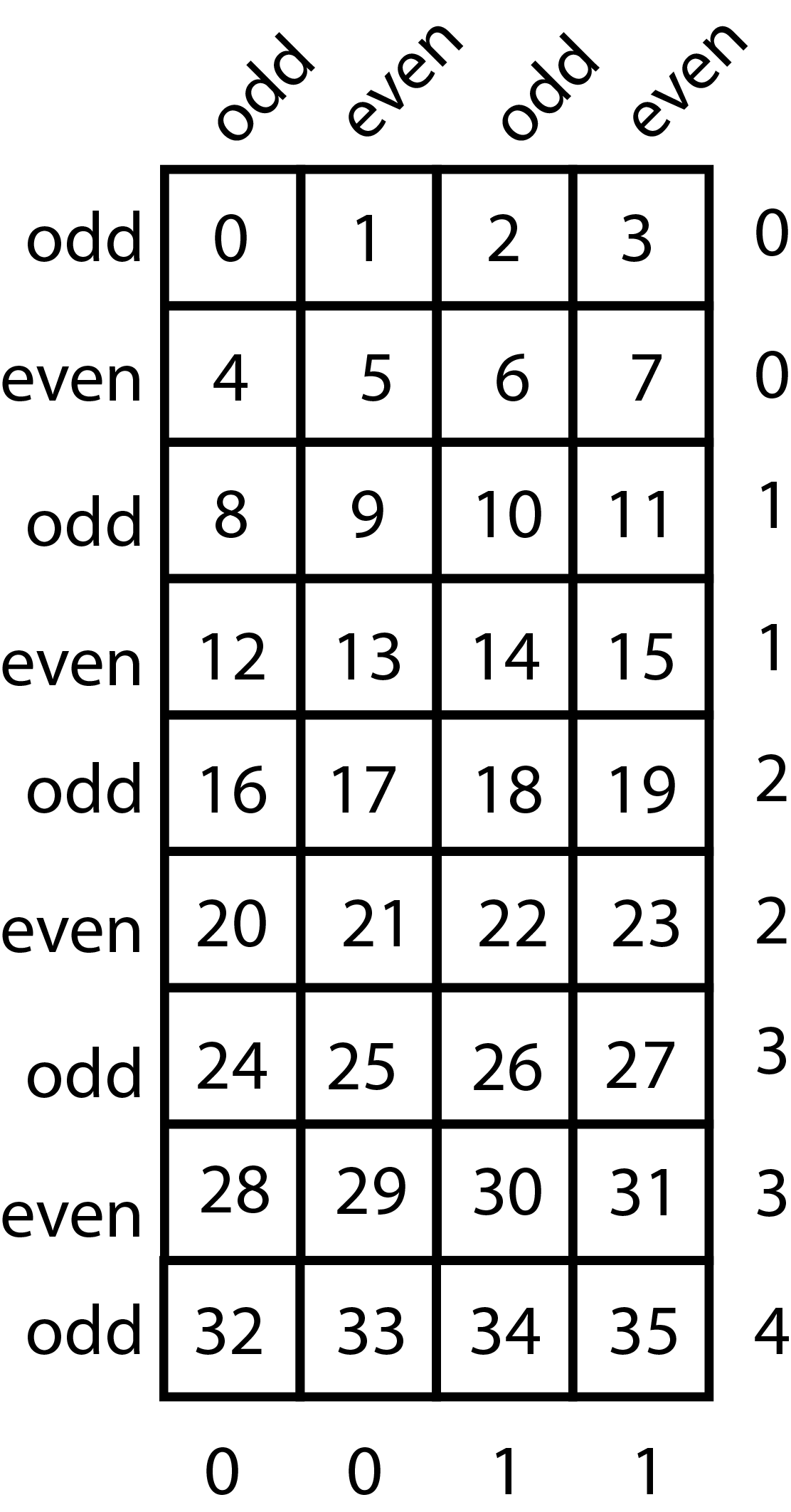}}
    \caption{New regions representation}
    \label{fig:select_regions}
\end{figure}

A new way to represent the selected regions by the row and column according to Figure~\ref{fig:select_regions} was suggested.

Every row and column was marked as either odd or even, and every odd-even pair was numbered. This way, the seed's neighborhood is defined by one even and one odd row and one even and one odd column. For example, the neighborhood including regions 4, 5, 8, and 9 is defined by even row 0 and odd row 1, even column 0 and odd column 0. Since the first row and the last row are both odd, two additional parameters were added to indicate whether the last row is used or not, and if it is used then whether it is considered even or odd. The grid represents an unfolded torus, and the first and the last rows are connected, e.g., region 0 has five adjacent regions: 1, 4, 5, 32, and 33.

It was thought that changes in the selected region’s representation should result in 8-to-1 MUXs (multiplexers) for row selection, 2-to-1 MUXs for last row selection, and 2-to-1 MUXs for column selection instead of the previous 36-to-1 MUXs. The problem was to make Vivado HLS infer it from the design.

A function that reads the data from the input buffer and writes it to the candidate arrays was redesigned in the following way. First, two rows should be selected: one even and one odd. Since the number of rows is uneven, the last row is considered as a special case. This way, if the candidate row is not the last one, then the selection is done from only four rows: 0, 2, 4, 6 for the even row and 1, 3, 5, and 7 for the odd. Then, from those two rows, two columns should be selected. Again, one even and one odd, therefore, in each case the selection is done from two columns: 0 and 2 for the even column and 1 and 3 for the odd.

In order to make Vivado~HLS synthesize appropriate MUXs, the following idea was implemented (presented in the example of tracks.) Tracks were saved in two arrays: 2D $4\times8$ track array for the first 8 rows and 1D trackLastRow array for the last row. Each row in the track array contains two rows from the original grid. This way, all even rows are on the left and all odd rows are on the right. Selected rows are copied to two 2D $2\times2$ arrays: tRowEven and tRowOdd. This way, even columns end up in the first column and odd columns in the second column.

\begin{figure}[h!]
    \centerline{\includegraphics[width=3.5in]{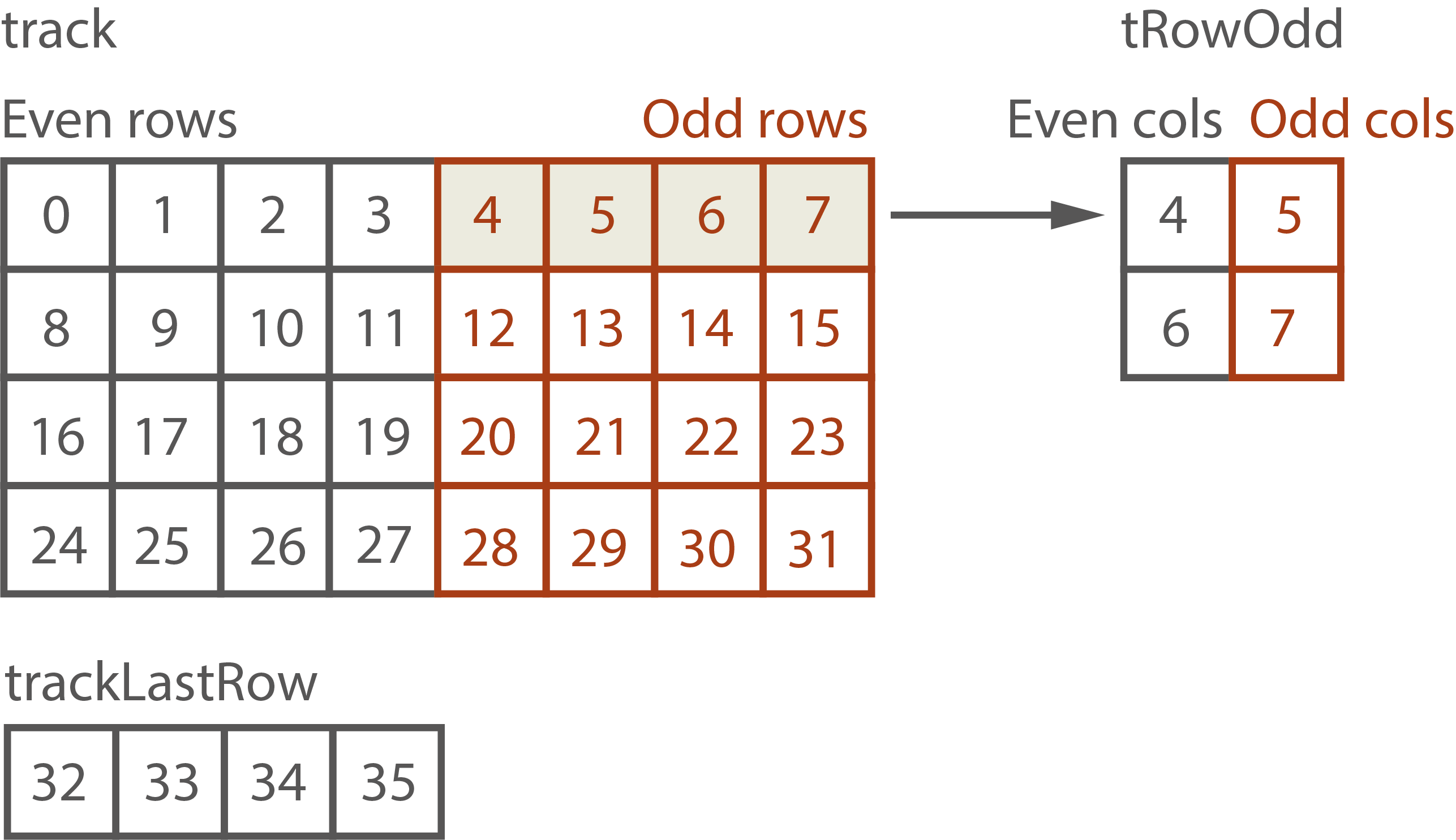}}
    \caption{Preselect candidates selection process}
    \label{fig:select_cands}
\end{figure}

Described architecture is presented in Figure~\ref{fig:select_cands}. Those shapes make it clear for Vivado~HLS which elements are valid for each selection. This design gives 2-to-1 MUXs for the last row selection, 4-to-1 MUXs for row selection, and 2-to-1 for column selection, which is even better than predicted.

This task made it clear that creating a well-optimized algorithm using HLS tools requires solving two problems: first, designing an efficient architecture and, second, writing an implementation that Vivado~HLS will be able to synthesize according to the idea.

\section{Conclusion}
For the first step of the algorithm, two different sorting algorithms were implemented and compared with the original solution: streaming merge sort and spatial insertion sort. The merge sort solution has shown that not every design can be implemented using HLS due to the limitations of the tools. It was not possible to write the code in a way that Vivado~HLS can schedule merge sorter with streaming data in parallel with the buffer process.

Insertion sort solution, on the other hand, was able to reduce the resource usage of the function drastically ($7\times$ fewer LUTs for the first version and $11\times$ fewer LUTs for the modified version compared to the original algorithm) as well as decrease the latency of the sorting part (21 \% for the modified version) and solve the problem with the routing. The suggested insertion sort algorithm considers the properties of the input data (new data comes presorted in blocks of 4) and the task specification (only 16 best elements out of 144 should be selected), proving that a solution tailored to the task provides a better design.

For the candidates selection task, a new way was introduced to select 4 regions for each seed and access the data from those regions in order to avoid huge multiplexers in the RTL design that originally caused problems with routing as well. The new design instead of 36-to-1 MUXs uses 4-to-1 and 2-to-1 MUXs, which take less area and do not cause problems with implementation on FPGA.

\bibliographystyle{IEEEtran}
\bibliography{literature}

\begin{thebibliography}{10}
\providecommand{\url}[1]{#1}
\csname url@samestyle\endcsname
\providecommand{\newblock}{\relax}
\providecommand{\bibinfo}[2]{#2}
\providecommand{\BIBentrySTDinterwordspacing}{\spaceskip=0pt\relax}
\providecommand{\BIBentryALTinterwordstretchfactor}{4}
\providecommand{\BIBentryALTinterwordspacing}{\spaceskip=\fontdimen2\font plus
\BIBentryALTinterwordstretchfactor\fontdimen3\font minus
  \fontdimen4\font\relax}
\providecommand{\BIBforeignlanguage}[2]{{%
\expandafter\ifx\csname l@#1\endcsname\relax
\typeout{** WARNING: IEEEtran.bst: No hyphenation pattern has been}%
\typeout{** loaded for the language `#1'. Using the pattern for}%
\typeout{** the default language instead.}%
\else
\language=\csname l@#1\endcsname
\fi
#2}}
\providecommand{\BIBdecl}{\relax}
\BIBdecl

\bibitem{Sakurai2014}
Y.~Sakurai, ``The {ATLAS} tau trigger performance during {LHC} {Run} 1 and
  prospects for {Run} 2,'' \emph{arXiv: High Energy Physics - Experiment}, Sep.
  2014.

\bibitem{Licht2021}
J.~de~Fine~Licht, M.~Besta, S.~Meierhans, and T.~Hoefler, ``Transformations of
  high-level synthesis codes for high-performance computing,'' \emph{{IEEE}
  Transactions on Parallel and Distributed Systems}, vol.~32, no.~5, pp.
  1014--1029, may 2021.

\bibitem{Matai2016}
J.~Matai, D.~Richmond, D.~Lee, Z.~Blair, Q.~Wu, A.~Abazari, and R.~Kastner,
  ``Resolve: Generation of high-performance sorting architectures from
  high-level synthesis,'' in \emph{Proceedings of the 2016 {ACM}/{SIGDA}
  International Symposium on Field-Programmable Gate Arrays}.\hskip 1em plus
  0.5em minus 0.4em\relax {ACM}, feb 2016.

\bibitem{Huang2020}
L.~Huang, D.-L. Li, K.-P. Wang, T.~Gao, and A.~Tavares, ``A survey on
  performance optimization of high-level synthesis tools,'' \emph{Journal of
  Computer Science and Technology}, vol.~35, no.~3, pp. 697--720, may 2020.

\bibitem{Liang2012}
Y.~Liang, K.~Rupnow, Y.~Li, D.~Min, M.~N. Do, and D.~Chen, ``High-level
  synthesis: Productivity, performance, and software constraints,''
  \emph{Journal of Electrical and Computer Engineering}, vol. 2012, pp. 1--14,
  2012.

\bibitem{Rupnow2011}
K.~Rupnow, Y.~Liang, Y.~Li, D.~Min, M.~Do, and D.~Chen, ``High level synthesis
  of stereo matching: Productivity, performance, and software constraints,'' in
  \emph{2011 International Conference on Field-Programmable Technology}.\hskip
  1em plus 0.5em minus 0.4em\relax {IEEE}, dec 2011.

\bibitem{Lahti2019}
S.~Lahti, P.~Sjovall, J.~Vanne, and T.~D. Hamalainen, ``Are we there yet? {A}
  study on the state of high-level synthesis,'' \emph{{IEEE} Transactions on
  Computer-Aided Design of Integrated Circuits and Systems}, vol.~38, no.~5,
  pp. 898--911, may 2019.

\bibitem{Cong2011a}
J.~Cong, B.~Liu, S.~Neuendorffer, J.~Noguera, K.~Vissers, and Z.~Zhang,
  ``High-level synthesis for {FPGAs}: From prototyping to deployment,''
  \emph{{IEEE} Transactions on Computer-Aided Design of Integrated Circuits and
  Systems}, vol.~30, no.~4, pp. 473--491, apr 2011.

\bibitem{ZhipengZhao2017}
Z.~Zhao and J.~C. Hoe, ``Using {Vivado HLS} for structural design,'' in
  \emph{Proceedings of the 2017 {ACM}/{SIGDA} International Symposium on
  Field-Programmable Gate Arrays}.\hskip 1em plus 0.5em minus 0.4em\relax
  {ACM}, feb 2017.

\bibitem{VivadoHLS}
\BIBentryALTinterwordspacing
Vivado design suite---{Vivado HLS}. Xilinx Inc. [Last access: 04.05.2021].
  [Online]. Available:
  \url{https://www.xilinx.com/products/design-tools/vivado.html}
\BIBentrySTDinterwordspacing

\bibitem{Kent2021}
R.~B. Kent and M.~S. Pattichis, ``Design, implementation, and analysis of
  high-speed single-stage {N}-sorters and {N}-filters,'' \emph{{IEEE} Access},
  vol.~9, pp. 2576--2591, 2021.

\end{thebibliography}

\end{document}